\documentclass[aps,prl,twocolumn,superscriptaddress]{revtex4-1}
\usepackage{graphicx}
\usepackage[hidelinks]{hyperref} 
\usepackage{bm}
\usepackage{amsmath}
\usepackage{dcolumn}% Align table columns on decimal point

\graphicspath{{/Users/shanggao/Documents/Paper/CESe_csi/PRL_round1_tom/}}

\begin{document}

% Use the \preprint command to place your local institutional report
% number in the upper righthand corner of the title page in preprint mode.
% Multiple \preprint commands are allowed.
% Use the 'preprintnumbers' class option to override journal defaults
% to display numbers if necessary
%\preprint{}

%Title of paper
\title{Dipolar spin ice states with fast monopole hopping rate in CdEr$\bf{_2}$\textit{X}$\bf{_4}$ (\textit{X} = Se, S)}

\author{Shang Gao}
%\email[]{Your e-mail address}
%\homepage[]{Your web page}
%\thanks{}
\affiliation{Laboratory for Neutron Scattering and Imaging, Paul Scherrer Institut, CH-5232 Villigen PSI, Switzerland}
\affiliation{Department of Quantum Matter Physics, University of Geneva, CH-1211 Geneva, Switzerland}

\author{O. Zaharko}
\email[]{oksana.zaharko@psi.ch}
%\homepage[]{Your web page}
%\thanks{}
\affiliation{Laboratory for Neutron Scattering and Imaging, Paul Scherrer Institut, CH-5232 Villigen PSI, Switzerland}

\author{V. Tsurkan}
%\email[]{Your e-mail address}
%\homepage[]{Your web page}
%\thanks{}
\affiliation{Experimental Physics V, University of Augsburg, D-86135 Augsburg, Germany}
\affiliation{Institute of Applied Physics, Academy of Sciences of Moldova, MD-2028 Chisinau, Republic of Moldova}

\author{L. Prodan}
%\email[]{Your e-mail address}
%\homepage[]{Your web page}
%\thanks{}
\affiliation{Institute of Applied Physics, Academy of Sciences of Moldova, MD-2028 Chisinau, Republic of Moldova}

\author{E. Riordan}
%\email[]{Your e-mail address}
%\homepage[]{Your web page}
%\thanks{}
\affiliation{School of Physics and Astronomy, Cardiff University, CF24 3AA Cardiff, United Kingdom}

\author{J. Lago}
%\email[]{Your e-mail address}
%\homepage[]{Your web page}
%\thanks{}
\affiliation{Department of Inorganic Chemistry, Universidad del Pa\'is Vasco (UPV-EHU), 48080 Bilbao, Spain}

\author{B. F{\aa}k}
%\email[]{Your e-mail address}
%\homepage[]{Your web page}
%\thanks{}
\affiliation{Institut Laue-Langevin, CS 20156, 38042 Grenoble Cedex 9, France}

\author{A. R. Wildes}
%\email[]{Your e-mail address}
%\homepage[]{Your web page}
%\thanks{}
\affiliation{Institut Laue-Langevin, CS 20156, 38042 Grenoble Cedex 9, France}

\author{M. M. Koza}
%\email[]{Your e-mail address}
%\homepage[]{Your web page}
%\thanks{}
\affiliation{Institut Laue-Langevin, CS 20156, 38042 Grenoble Cedex 9, France}

\author{C. Ritter}
%\email[]{Your e-mail address}
%\homepage[]{Your web page}
%\thanks{}
\affiliation{Institut Laue-Langevin, CS 20156, 38042 Grenoble Cedex 9, France}

\author{P. Fouquet}
%\email[]{Your e-mail address}
%\homepage[]{Your web page}
%\thanks{}
\affiliation{Institut Laue-Langevin, CS 20156, 38042 Grenoble Cedex 9, France}

\author{L. Keller}
%\email[]{Your e-mail address}
%\homepage[]{Your web page}
%\thanks{}
\affiliation{Laboratory for Neutron Scattering and Imaging, Paul Scherrer Institut, CH-5232 Villigen PSI, Switzerland}

\author{E. Can{\'e}vet}
%\email[]{Your e-mail address}
%\homepage[]{Your web page}
%\thanks{}
\affiliation{Laboratory for Neutron Scattering and Imaging, Paul Scherrer Institut, CH-5232 Villigen PSI, Switzerland}
\affiliation{Department of Physics, Technical University of Denmark, DK-2800 Kgs. Lyngby, Denmark}

\author{M. Medarde}
%\email[]{Your e-mail address}
%\homepage[]{Your web page}
%\thanks{}
\affiliation{Laboratory for Scientific Developments and Novel Materials, Paul Scherrer Institut, CH-5232 Villigen PSI, Switzerland}

\author{J. Blomgren}
%\email[]{Your e-mail address}
%\homepage[]{Your web page}
%\thanks{}
\affiliation{RISE Acreo AB, SE-411 33 G\"oteborg, Sweden}

\author{C.~Johansson}
%\email[]{Your e-mail address}
%\homepage[]{Your web page}
%\thanks{}
\affiliation{RISE Acreo AB, SE-411 33 G\"oteborg, Sweden}

\author{S. R. Giblin}
%\email[]{Your e-mail address}
%\homepage[]{Your web page}
%\thanks{}
\affiliation{School of Physics and Astronomy, Cardiff University, CF24 3AA Cardiff, United Kingdom}

\author{S. Vrtnik}
%\email[]{Your e-mail address}
%\homepage[]{Your web page}
%\thanks{}
\affiliation{Jo{\v z}ef Stefan Institute, SI-1000 Ljubljana, Slovenia}

\author{J. Luzar}
%\email[]{Your e-mail address}
%\homepage[]{Your web page}
%\thanks{}
\affiliation{Jo{\v z}ef Stefan Institute, SI-1000 Ljubljana, Slovenia}

\author{A. Loidl}
%\email[]{Your e-mail address}
%\homepage[]{Your web page}
%\thanks{}
\affiliation{Experimental Physics V, University of Augsburg, D-86135 Augsburg, Germany}

\author{Ch. R\"uegg}
%\email[]{Your e-mail address}
%\homepage[]{Your web page}
%\thanks{}
\affiliation{Laboratory for Neutron Scattering and Imaging, Paul Scherrer Institut, CH-5232 Villigen PSI, Switzerland}
\affiliation{Department of Quantum Matter Physics, University of Geneva, CH-1211 Geneva, Switzerland}

\author{T. Fennell}
\email[]{tom.fennell@psi.ch}
%\homepage[]{Your web page}
%\thanks{}
\affiliation{Laboratory for Neutron Scattering and Imaging, Paul Scherrer Institut, CH-5232 Villigen PSI, Switzerland}

%Collaboration name if desired (requires use of superscriptaddress
%option in \documentclass). \noaffiliation is required (may also be
%used with the \author command).
%\collaboration can be followed by \email, \homepage, \thanks as well.
%\collaboration{}
%\noaffiliation

\date{\today}

\begin{abstract}
Excitations in a spin ice behave as magnetic monopoles, and their population and mobility control the dynamics of a spin ice at low temperature.  CdEr$_2$Se$_4$ is reported to have the Pauling entropy characteristic of a spin ice, but its dynamics are three-orders of magnitude faster than the canonical spin ice Dy$_2$Ti$_2$O$_7$. In this letter we use diffuse neutron scattering to show that both CdEr$_2$Se$_4$ and CdEr$_2$S$_4$ support a dipolar spin ice state -- the host phase for a Coulomb gas of emergent magnetic monopoles.  These Coulomb gases have similar parameters to that in Dy$_2$Ti$_2$O$_7$, \textit{i.e.} dilute and uncorrelated, so cannot provide three-orders faster dynamics through a larger monopole population alone.  We investigate the monopole dynamics using ac susceptometry and neutron spin echo spectroscopy, and verify the crystal electric field Hamiltonian of the Er$^{3+}$ ions using inelastic neutron scattering.  A quantitative calculation of the monopole hopping rate using our Coulomb gas and crystal electric field parameters shows that the fast dynamics in CdEr$_2X_4$ ($X$ = Se, S) are primarily due to much faster monopole hopping.  Our work suggests that CdEr$_2X_4$ offer the possibility to study alternative spin ice ground states and dynamics, with equilibration possible at much lower temperatures than the rare earth pyrochlore examples.

\end{abstract}

% insert suggested PACS numbers in braces on next line
% insert suggested keywords - APS authors don't need to do this
%\keywords{}

%\maketitle must follow title, authors, abstract, \pacs, and \keywords
\maketitle

% body of paper here - Use proper section commands
% References should be done using the \cite, \ref, and \label commands
% \section{I. Introduction}

A magnetic Coulomb phase is characterized by an effective magnetic field whose topological defects behave as emergent magnetic monopoles~\cite{henley_coulomb_2010}.  In dipolar spin ices such as Dy$_2$Ti$_2$O$_7$, where long-range dipolar interactions between spins on the pyrochlore lattice establish the two-in-two-out ice rule (which gives the field its non-divergent character)~\cite{bramwell_spin_2001}, the monopoles are deconfined and interact according to a magnetic Coulomb law~\cite{castelnovo_magnetic_2008, fennell_magnetic_2009, morris_dirac_2009}. The transformation from the spin model to a Coulomb gas of magnetic monopoles simplifies the understanding of the properties of dipolar spin ices as the complicated couplings among the spins are replaced by the determinant parameters of the Coulomb gas: the elementary charge $Q_m$, chemical potential $v_0$, and hopping rate $u$~\cite{castelnovo_magnetic_2008, giblin_creation_2011}. Through analogs with Debye-H\"uckel theory of Coulomb gases, many thermodynamic observables can be conveniently calculated~\cite{ryzhkin_magnetic_2005,castelnovo_debye_2011, zhou_high_2011}.

The spin relaxation rate of canonical spin ices was a particular problem in the spin representation. From high to low temperature it changes from thermally activated, to a temperature independent plateau, to a re-entrant thermally activated regime \cite{snyder_low_2004, jaubert_signature_2009, jaubert_magnetic_2011, ruminy_phonon_2017}. At high temperature, above the monopole regime, Orbach processes describe the thermally activated relaxation rate~\cite{ruminy_phonon_2017}. The plateau and re-entrant thermally activated regimes are not readily explained in the spin representation, but can now be understood as the hopping of monopoles by quantum tunneling in screened and unscreened regimes of the Coulomb gas respectively~\cite{jaubert_signature_2009, jaubert_magnetic_2011}. In the unscreened regime, the relaxation rate depends on the monopole density $\rho$ with the hopping rate $u$ as the coefficient: $f\propto u\rho$ when the system is near equilibrium~\cite{ryzhkin_magnetic_2005, castelnovo_debye_2011, paulsen_far_2014}.

Although the monopole charge $Q_m$ and chemical potential $v_0$ can be calculated exactly from the spin model, the value of the monopole hopping rate $u$ is not well-understood and is usually treated as a fitting parameter~\cite{jaubert_signature_2009,jaubert_magnetic_2011, takatsu_ac_2013}. For Dy$_2$Ti$_2$O$_7$, $u$ is fitted to be $\sim 10^{3}$~Hz at $T< 12$~K, which has been experimentally confirmed through the Wien effect \cite{giblin_creation_2011}. Recently, Tomasello \textit{et al} found that this hopping rate can be estimated by the splitting of the crystal-electric-field (CEF) ground state doublet under an internal transverse magnetic field of 0.1--1~T~\cite{tomasello_single_2015}. To verify the universality of this approach, it is beneficial to compare the monopole dynamics in other dipolar spin ice compounds.

The newly proposed spin ice state in the spinel CdEr$_2$Se$_4$ provides such an opportunity~\cite{lau_geometrical_2005, lago_cder2se4_2010, plessis_spin_2017}. In this compound, Er$^{3+}$ ions constitute the pyrochlore lattice, and bulk measurements have revealed the Pauling entropy and local Ising character for the Er$^{3+}$ spins~\cite{lau_geometrical_2005, lago_cder2se4_2010}; both are strong indicators of the existence of the spin ice state although microscopic evidence is required to confirm the dipolar character necessary for deconfined, interacting monopoles. Of special importance is the low-temperature dynamics in CdEr$_2$Se$_4$, which was revealed to be three-orders faster than that of the pyrochlore titanate Dy$_2$Ti$_2$O$_7$~\cite{lago_cder2se4_2010}. The origin of this increase and its compatibility with monopole dynamics in CdEr$_2$Se$_4$ remains unclear.

In this letter, we explore spin ice states and monopole dynamics in CdEr$_2X_4$ ($X$ = Se, S). Using inelastic neutron scattering to study the CEF transitions and neutron diffuse scattering to study the spin correlations, we confirm the existence of dipolar spin ice states in CdEr$_2X_4$. Through ac susceptibility measurements, we reveal fast monopole dynamics in the whole quantum tunneling regime. Comparison with a calculation of the splitting of the Er$^{3+}$ CEF ground state doublet under perturbative transverse fields reveals the increase of the monopole hopping rate as the main contribution to the fast dynamics. Thus our work explains the fast monopole dynamics in CdEr$_2X_4$ and provides general support to this monopole hopping mechanism in dipolar spin ices. 

Our powder samples of CdEr$_2$Se$_4$ and CdEr$_2$S$_4$ were synthesized by the solid state reaction method~\cite{supp}. To reduce neutron absorption, the $^{114}$Cd isotope was used. X-ray diffraction measurements confirmed the good quality of our samples, with the Er$_xX_y$ impurities less than 1~\%. Inelastic neutron scattering experiments were performed on IN4 with 1.21 and 2.41~\AA\ incident neutron wavelengths at Institut Laue-Langevin (ILL). Polarized neutron diffuse scattering experiments were performed on CdEr$_2$Se$_4$ using D7 with a 4.8~\AA\ setup at ILL. Non-polarized neutron diffuse scattering experiments were performed on CdEr$_2$S$_4$ using DMC with a 2.46~\AA\ setup at SINQ of Paul Scherrer Insitut (PSI). Neutron spin echo experiments were performed on IN11 at ILL. AC susceptibilities $\chi$ in the frequency range of $1$--$1\times 10^3$~Hz were measured with the Quantum Design MPMS SQUID at Laboratory for Scientific Developments and Novel Materials of PSI. AC susceptibilities in the frequency range of $2.5\times 10^4$--$5.5\times 10^6$~Hz were measured using a bespoke induction ac susceptometer.

%---------------------------------------------------------
\begin{figure}
\includegraphics[width=0.46\textwidth]{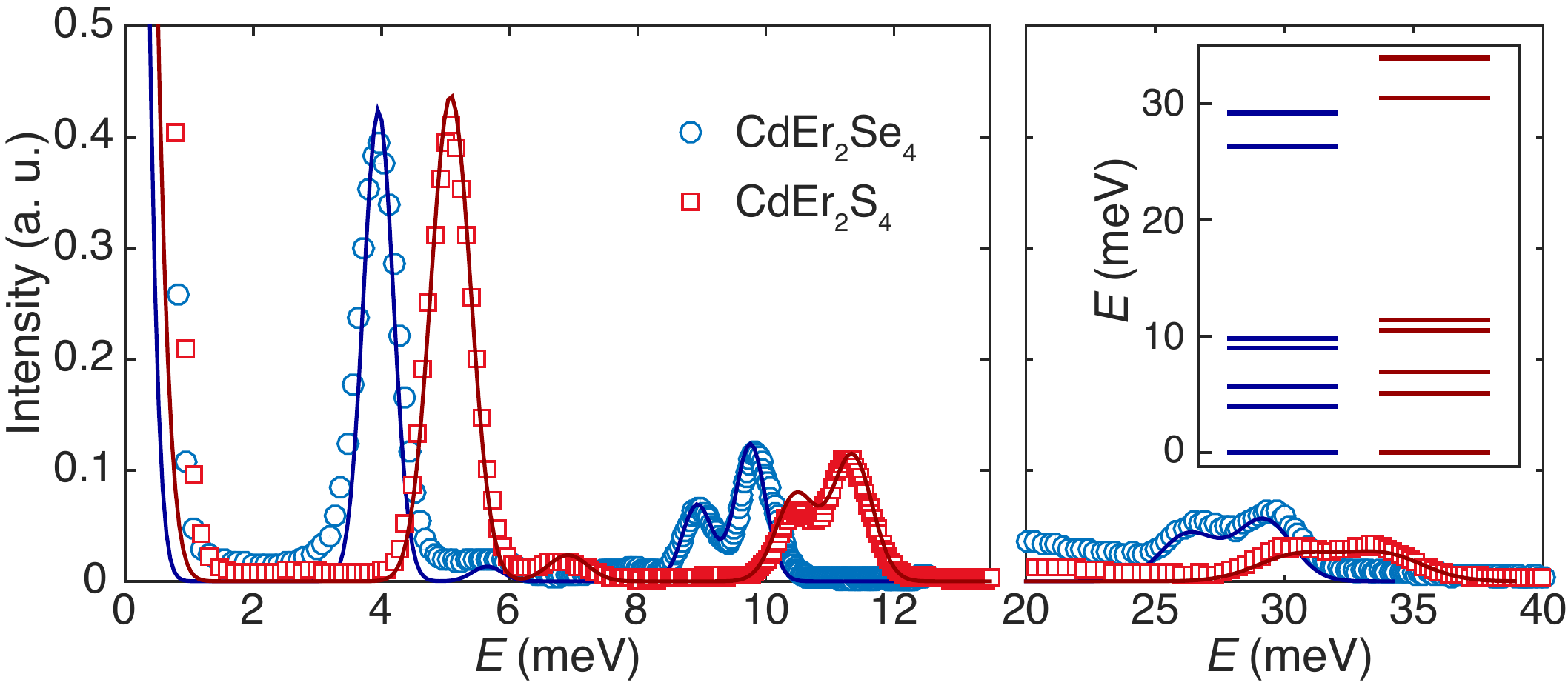}
\caption{(color online). Inelastic neutron scattering results of the CEF transitions in CdEr$_2$Se$_4$ (measured at $T = 2$~K) and CdEr$_2$S$_4$ (measured at $T=1.5$~K). Error bars are smaller than the symbol size. The fits are shown as the solid lines. The inset shows the fitted energies of the CEF levels for CdEr$_2$Se$_4$ (left column) and CdEr$_2$S$_4$ (right column).
}
\label{fig:cef}
\end{figure}
%---------------------------------------------------------

Fig.~\ref{fig:cef} presents the inelastic neutron scattering results of the CEF transitions in CdEr$_2$Se$_4$ and CdEr$_2$S$_4$. Altogether 6 peaks are observed at the base temperature for both compounds, which is consistent with the Stokes transitions within the Er$^{3+}$ $^{4}I_{15/2}$ manifold under $D_{3d}$ symmetry. Using the McPhase program \cite{rotter_using_2004}, we fitted the measured spectra with the CEF Hamiltonian $\mathcal{H} = \sum_{lm}B_l^m\hat{O}_l^m$, where $\hat{O}_l^m$ are the Stevens operators and $B_l^m$ are the corresponding coefficients. The fitting results are shown in Fig.~\ref{fig:cef} as the solid lines and Table~\ref{tab:cef} lists the fitted CEF parameters and ground state wavefunctions.  The energies of the CEF levels are presented in the inset of Fig.~\ref{fig:cef}, and also in the supplemental materials~\cite{supp}. For both compounds, the ground states transform as the $\Gamma_5^{+}\oplus \Gamma_6^{+}$ dipole-octupole doublet~\cite{huang_quantum_2014, li_hidden_2016}. Specifically, the wavefunctions for both of the ground state doublets are dominated by the $|15/2, \pm15/2\rangle$ components and have almost the same anisotropic $g$-factors of $g_{\perp} = 0$ and $g_{\parallel} = 16.4$, which is consistent with the previous report for CdEr$_2$Se$_4$~\cite{lago_cder2se4_2010}. Thus our inelastic neutron scattering results confirm the Ising character of the Er$^{3+}$ spins in CdEr$_2$Se$_4$ and CdEr$_2$S$_4$.  Scaling our parameters~\cite{supp} suggests other members of the series may be: Heisenberg-like (Dy, Yb); non-magnetic (Tm); or Ising-like with low-lying excited states (Ho), a situation of interest for forming a quantum spin ice~\cite{molavian_dynamically_2007}

%--------------------------------------------------------
\begin{table}[t]
\caption{The fitted Wybourne CEF parameters (meV) and ground state doublets for CdEr$_2$Se$_4$ and CdEr$_2$S$_4$.}
\label{tab:cef}
\centering
\begin{tabular}{lcccccc}
\toprule
 & $B_2^0$ & $B_4^0$ & $B_4^3$ & $B_6^0$ & $B_6^3$ & $B_6^6$ \\
 \hline
CdEr$_2$Se$_4$	& $-25.70$& $-107.73$ & $-97.74\, $  & $25.31$ & $-19.06$ & $9.51$\\
CdEr$_2$S$_4$	& $-29.18$ & $-122.72$ & $-113.66\, $ & $25.97$ & $-21.89$ & $14.41$\\
\hline
\end{tabular}
\begin{tabular}{lccccc}
 \multicolumn{1}{c}{$J_z$} & $\pm15/2$ & $\pm 9/2$ & $\pm 3/2$ & $\mp 3/2$ & $\mp 9/2$ \\
 \hline
CdEr$_2$Se$_4\ $& $\ \pm 0.906\ $ & $\ \, 0.386\ $ & $\ \pm 0.159\ $ & $\ -0.073\ $ & $\ \pm 0.004\ $ \\
CdEr$_2$S$_4\ $& $\ \pm 0.904\ $ & $\ \, 0.391\ $ & $\ \pm 0.145\ $ & $\ -0.094\ $ & $\ \pm 0.006\ $ \\
\botrule
\end{tabular}
\end{table}
%--------------------------------------------------------

\begin{figure} [t]
\includegraphics[width=0.435\textwidth]{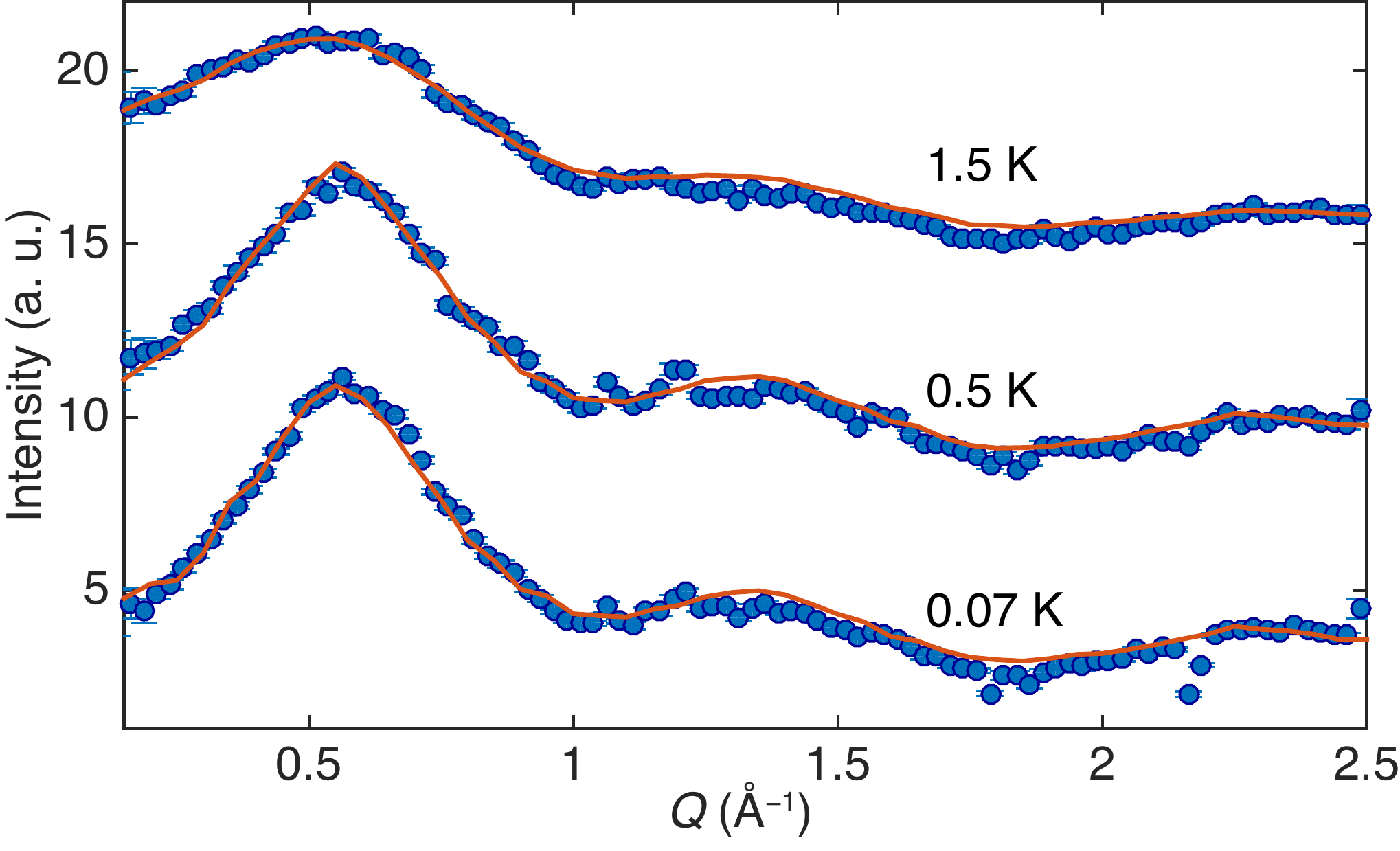}
\caption{(color online). CdEr$_2$Se$_4$ magnetic scattering at 0.07, 0.5, and 1.5 K obtained from the $xyz$ polarization analysis. The 0.5 (1.5) K data is shifted by 6 (12) along the $y$ axis. The Monte Carlo simulation results are shown as the solid red lines.}
\label{fig:diff}
\end{figure}

Although the Pauling entropy is a strong signature of the spin ice state in CdEr$_2$Se$_4$~\cite{lago_cder2se4_2010}, it only characterizes the spin configurations at the length scale of a single tetrahedron. To realize a magnetic Coulomb gas with interacting monopoles, it is essential to have a dipolar spin ice state with power-law spin correlations, which can be verified through measurements of the spin correlations \cite{fennell_magnetic_2009}. Fig.~\ref{fig:diff} presents the quasi-static spin-spin correlations in CdEr$_2$Se$_4$ obtained from polarized neutron diffuse scattering \cite{ehlers_generalization_2013}. Broad peaks are observed at 0.6 and 1.4~\AA$^{-1}$, and the overall pattern is very similar to that of the known dipolar spin ices~\cite{kadowaki_neutron_2002,mirebeau_spin_2004,hallas_statics_2012}. Sharp peaks with very weak intensities are also discernible near 1.1~\AA$^{-1}$ and can be attributed to the magnetic Bragg peaks of Er$_x$Se$_y$ impurities \cite{supp}.

To fit the observed spin-spin correlations in CdEr$_2$Se$_4$, we performed single-spin-flip Monte Carlo simulations for the dipolar spin ice model with exchange couplings up to the second neighbors \cite{yavorskii_dy2ti2o7_2008}:
\begin{eqnarray}
\label{eq:dsi}
\mathcal{H} &=& J_1\sum_{\langle ij \rangle} \sigma_i \sigma_j + J_2\sum_{\langle\langle ij \rangle\rangle} \sigma_i \sigma_j \notag \\
&+& Dr_0^3\sum_{ij} \Bigg[ \frac{\vec{n}_i\cdot \vec{n}_j}{|r_{ij}|^3} - \frac{3(\vec{n}_i\cdot \vec{r}_{ij})(\vec{n}_j\cdot \vec{r}_{ij})}{|r_{ij}|^5} \Bigg] \sigma_i \sigma_j \text{.}
\end{eqnarray}
Here, $\vec{n}_i$ is the unit vector along the local $\langle 111\rangle$ axes with the positive direction pointing from one diamond sublattice of the tetrahedra center to the other, $\sigma_i = \pm 1$ is the corresponding Ising variable, $J_1$ and $J_2$ are the exchange interactions for nearest neighbors (NN) $\langle ij \rangle$ and second-nearest neighbors $\langle\langle ij \rangle\rangle$, respectively, $r_0$ is the NN distance, and $D = \mu_0 (\langle \hat{J}_z \rangle g\mu_B)^2/(4\pi r_0^3)$ is the dipolar interaction, 0.62 and 0.69~K for CdEr$_2$Se$_4$ and CdEr$_2$S$_4$, respectively. With the ALPS package \cite{bauer_alps_2011}, we implemented the Hamiltonian~(\ref{eq:dsi}) on a $6\times 6\times 6$ supercell with periodic boundary conditions. The dipolar interaction was truncated beyond the distance of 3 unit cells. The spin-spin correlations were evaluated every 100 sweeps during the $4\times 10^5$ sweeps of measurement. Assuming the effective NN coupling $J_{\textrm{eff}} = J_1 + 5D/3$ to be equal to 1 K at which temperature the CdEr$_2$Se$_4$ specific heat maximum was observed~\cite{lago_cder2se4_2010, hertog_dipolar_2000, zhou_chemical_2012}, we fixed $J_1$ to $-0.03(1)$~K and only varied $J_2$ in the fitting process. As is shown in Fig.~\ref{fig:diff}, the model with $J_2=0.04(1)$~K fits the measured spin correlations very well. We found no need to include $J_3$, which appears in  other dipolar spin ices~\cite{yavorskii_dy2ti2o7_2008}. Although the exact value of $J_2$ might be susceptible to both the supercell size and the dipolar cutoff, our simulations do confirm the dominance of the dipolar interactions in CdEr$_2$Se$_4$. Non-polarized neutron diffuse scattering results for CdEr$_2$S$_4$ are shown in the Supplemental Material \cite{supp}, which have similar $Q$-dependence as that of CdEr$_2$Se$_4$ and can be fitted by the dipolar spin ice model as well. In this way, we establish the existence of the dipolar spin ice state in CdEr$_2$Se$_4$ and CdEr$_2$S$_4$.

%--------------------------------------------------------
%--------------------------------------------------------

With the fitted CEF ground states and coupling strengths, we can determine the monopole parameters. The monopole charge $Q_m = 2\langle \hat{J}_z \rangle g \mu_B / \sqrt{3/2} r_0$ can be calculated to be 3.28 and 3.42~$\mu_B/$\AA\ for CdEr$_2$Se$_4$ and CdEr$_2$S$_4$, respectively~\cite{castelnovo_magnetic_2008}. The chemical potential $v_0 = 2J_1+(8/3)(1+\sqrt{2/3})D$, which is half of the energy cost to create and unbind a monopole-antimonopole pair~\cite{zhou_high_2011}, is 2.93~K for CdEr$_2$Se$_4$ and 3.84~K for CdEr$_2$S$_4$. Although the chemical potentials in CdEr$_2X_4$ are lower than that in Dy$_2$Ti$_2$O$_7$ (4.35 K), they are still more than two times higher than the energy cost $E_{\textrm{unbind}} = (8/3)\sqrt{2/3}D$ to unbind a monopole-antimonopole pair, locating both compounds in the same weakly correlated magnetolyte regime as Dy$_2$Ti$_2$O$_7$~\cite{zhou_high_2011}.

%---------------------------------------------------------
\begin{figure} [t]
\includegraphics[width=0.46\textwidth]{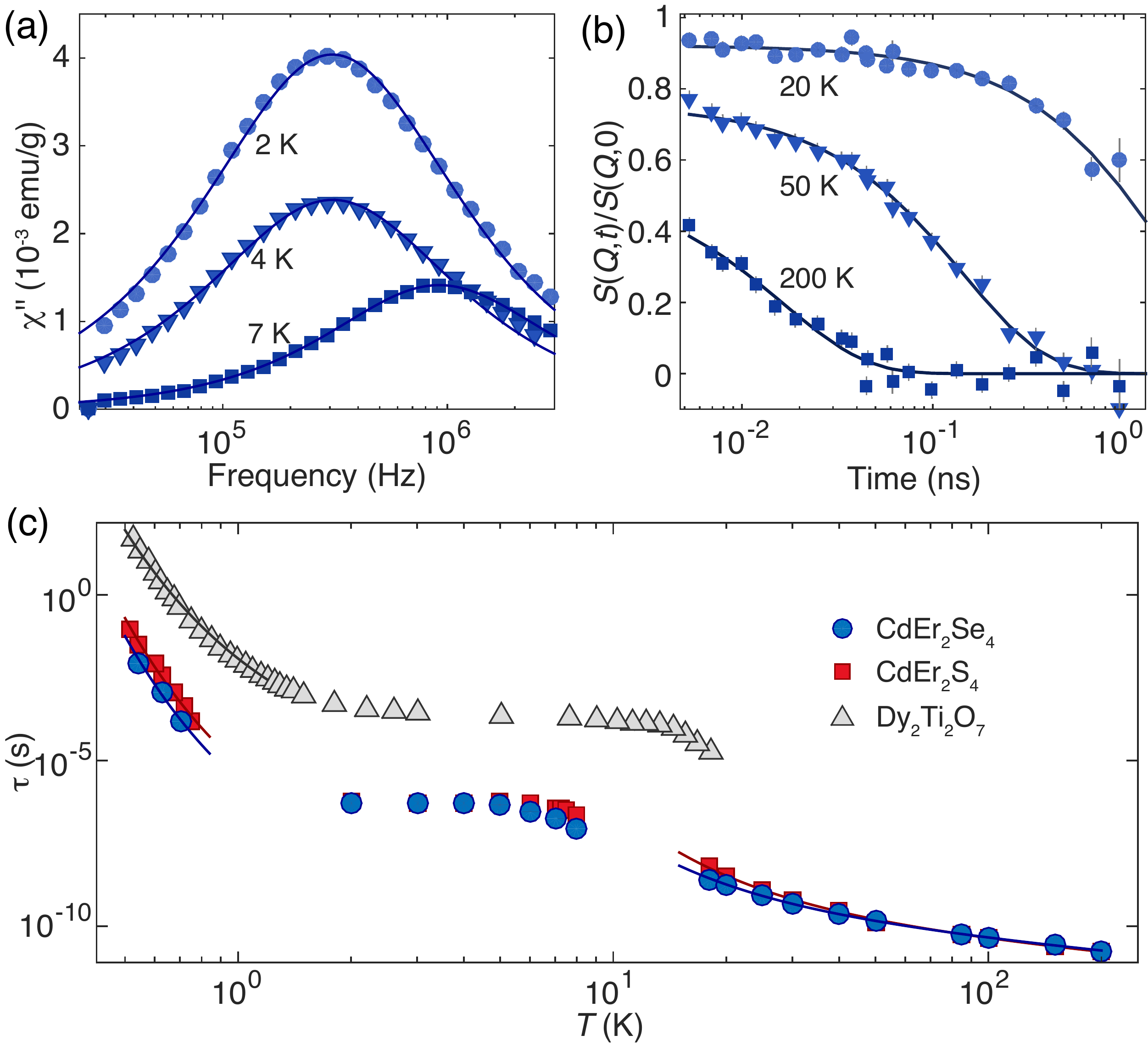}
\caption{(color online). (a) Imaginary parts of the ac susceptibilities of CdEr$_2$Se$_4$ measured at 2, 4, and 7 K with the Cole-Cole model fits shown as the solid lines. (b) Normalized spin echo intermediate scattering function $S(Q,t)/S(Q,0)$ of CdEr$_2$Se$_4$ measured at 20, 50, and 200 K with the fits shown as the solid lines. (c) Extracted relaxation time in CdEr$_2$Se$_4$ and CdEr$_2$S$_4$. Error bars are smaller than the symbol sizes. The Arrhenius (Orbach) fits in the low (high) temperature regime are shown as the solid lines. Relaxation rates together with the low-temperature Arrhenius fits for Dy$_2$Ti$_2$O$_7$~\cite{yaraskavitch_spin_2012, matsuhira_novel_2001} are shown for comparison.}
 \label{fig:mono_obs}
\end{figure}
%---------------------------------------------------------

Monopole dynamics in the low and high frequency regimes can be probed with ac-susceptibility~\cite{snyder_low_2004, quilliam_dynamics_2011, yaraskavitch_spin_2012, matsuhira_novel_2001} and neutron spin echo spectroscopy~\cite{ehlers_dynamical_2003, ehlers_evidence_2004}, respectively, and the representative results for CdEr$_2$Se$_4$ are shown in Fig.~\ref{fig:mono_obs}a and b. Fig.~\ref{fig:mono_obs}c summarizes the temperature dependence of the characteristic relaxation time $\tau = 1/2\pi f$ in CdEr$_2X_4$, where the results for $\tau>1\times 10^{-3}$~s are extracted from the peak positions of the imaginary part of the ac-susceptibility $\chi''(T)$, the results for $10^{-5}>\tau>10^{-7}$~s are  obtained by fitting $\chi(\omega)$ to the Cole-Cole model \cite{bovo_brownian_2013},  and the results for $\tau<10^{-8}$ s are obtained by fitting the neutron spin echo intermediate scattering function with $S(Q,t)/S(Q,0)= A\operatorname{exp}[-t/\tau(T)]$~\cite{ehlers_dynamical_2003, ehlers_evidence_2004}. The relaxation time in Dy$_2$Ti$_2$O$_7$~\cite{yaraskavitch_spin_2012, matsuhira_novel_2001} is also shown in Fig.~\ref{fig:mono_obs}c for comparison.

Firstly, we observe that at $T>10$ K, the relaxation time in CdEr$_2X_4$ obeys the Orbach law of $\tau = \tau_0[\operatorname{exp}(\Delta/k_BT)-1]$~\cite{ruminy_phonon_2017}, with the parameters $\tau_0 = 3.93(9)\times 10^{-11}$~s and $\Delta = 77.1$~K for CdEr$_2$Se$_4$, and $\tau_0 = 2.73(5)\times 10^{-11}$~s and $\Delta = 96.3$~K for CdEr$_2$S$_4$. The fitted excitation energies $\Delta$ in CdEr$_2X_4$ are much smaller than that of Dy$_2$Ti$_2$O$_7$ ($\Delta > 230$ K), which is due to their lower CEF excited states~\cite{ruminy_phonon_2017}.

The Orbach behavior of the relaxation rate does not extend to the lowest temperature. Instead, at $T$ in-between 2 and 5 K, a plateau region with $\tau \sim 4.9 \times 10^{-7}$~s, which was inaccessible in the previous susceptibility measurements~\cite{lago_cder2se4_2010}, is observed for both CdEr$_2$Se$_4$ and CdEr$_2$S$_4$, reminiscent of the $\tau \sim 2.6 \times 10^{-4}$~s quantum tunneling plateau in Dy$_2$Ti$_2$O$_7$~\cite{snyder_low_2004, jaubert_signature_2009, matsuhira_novel_2001}. Such a similarity extends to even lower temperatures where the relaxation time starts rising again. As can be seen in Fig.~\ref{fig:mono_obs}c, at $T < 1$ K, the relaxation time in CdEr$_2X_4$ can be described by the Arrhenius law of $\tau_0\operatorname{exp}(\Delta/k_BT)$, with parameters $\tau_0 = 1.01(1) \times 10^{-10}$~s and $\Delta = 10.07$~K for CdEr$_2$Se$_4$, and $\tau_0 = 2.9(1) \times 10^{-10}$~s and $\Delta = 10.2(6)$~K for CdEr$_2$S$_4$. Due to the limited data points for CdEr$_2$Se$_4$, the fitted $\Delta$ value from Ref.~\cite{lago_cder2se4_2010} has been used. The activation energies in CdEr$_2X_4$ are very close to that of Dy$_2$Ti$_2$O$_7$, where the Arrhenius law with $\tau_0 = 3.07 \times 10^{-7}$~s and $\Delta = 9.93$~K has been observed in a similar temperature regime~\cite{yaraskavitch_spin_2012}.

Despite the similar temperature evolution, the absolute values of the monopole relaxation rates in CdEr$_2X_4$ are about $10^3$ times higher than that in Dy$_2$Ti$_2$O$_7$ for the whole measured quantum tunneling region, which cannot be simply accounted for by the difference of the monopole densities $\rho$. Assuming $\rho(T) \propto \operatorname{exp}(-v_0/k_BT) $, the monopole densities in CdEr$_2X_4$ are no more than $10$ times higher than that of Dy$_2$Ti$_2$O$_7$ in the investigated quantum tunneling region. According to the $f\propto u\rho$ relation of the Debye-H\"uckel theory, there must be a two-orders increase of the monopole hopping rates $u$ in CdEr$_2X_4$.

%---------------------------------------------------------
\begin{figure} [t]
\includegraphics[width=0.44\textwidth]{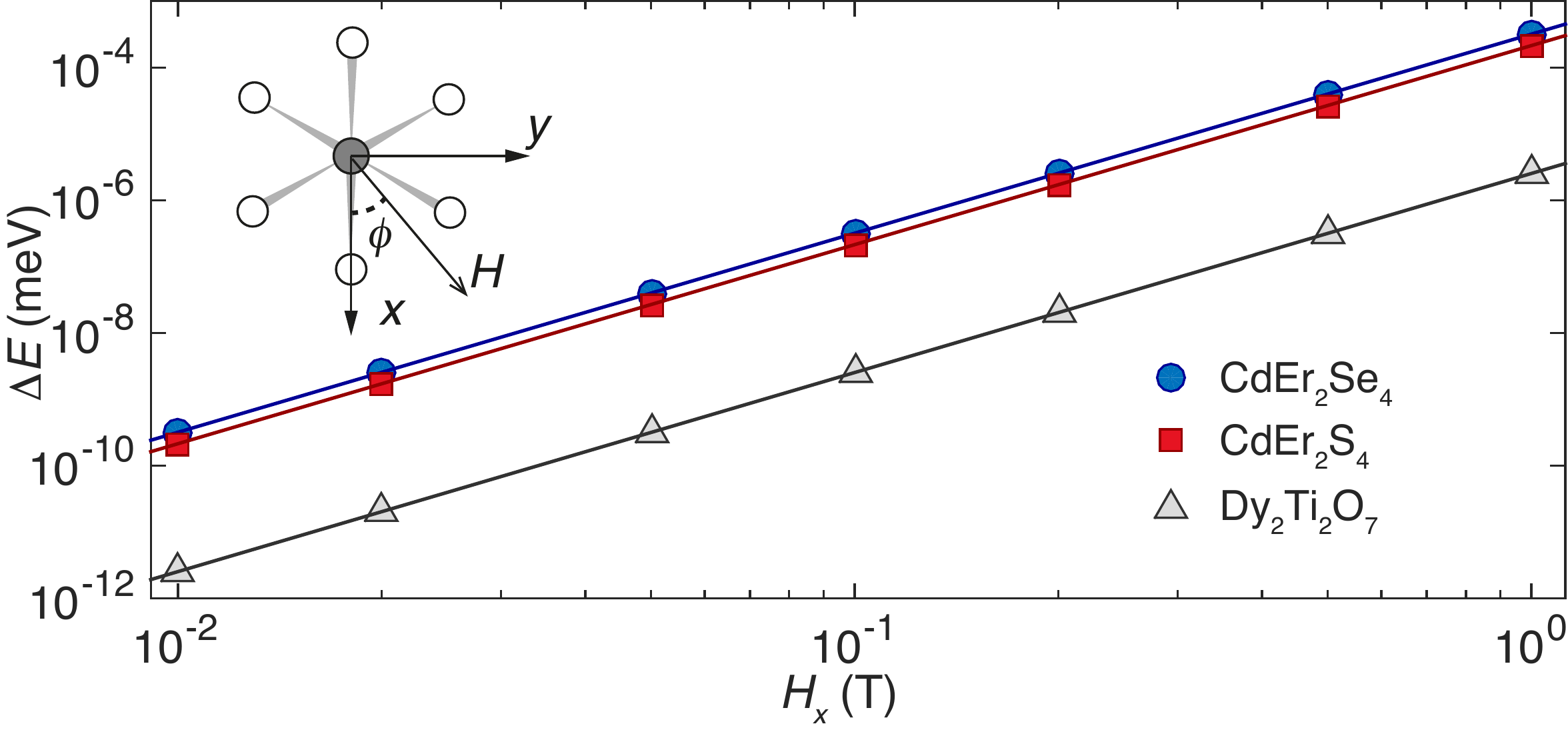}
\caption{(color online). Splittings of the CEF ground state doublet in CdEr$_2$Se$_4$, CdEr$_2$S$_4$, and Dy$_2$Ti$_2$O$_7$ under a perturbative magnetic field along the $x$ direction. Definitions of the axes are shown in the inset.}
\label{fig:mono_cal}
\end{figure}
%---------------------------------------------------------

Following Tomasello \textit{et al.} \cite{tomasello_single_2015}, we analyze the perturbation effect of an internal transverse magnetic field on the CEF ground state doublet in CdEr$_2$Se$_4$ and CdEr$_2$S$_4$. Due to the similar NN couplings~\cite{zhou_high_2011}, we expect similar internal field strengths in CdEr$_2X_4$ and Dy$_2$Ti$_2$O$_7$~\cite{sala_magnetic_2012}. The perturbed Hamiltonian can be written as:
\begin{equation}
\label{eq:stevens_perturb}
\mathcal{H} = \sum_{lm} B_l^m \hat{O}_l^m + H\operatorname{cos}(\phi) \hat{J}_x + H\operatorname{sin}(\phi) \hat{J}_y \text{,} 
\end{equation}
where the $y$ direction is along the $C_2$ axis and $\phi$ is the angle between the transverse field $H$ and the $x$ direction (see the inset of Fig.~\ref{fig:mono_cal}). Similar to the Dy$^{3+}$ ions in Dy$_2$Ti$_2$O$_7$~\cite{tomasello_single_2015}, the Kramers degeneracy of the Er$^{3+}$ ions causes a third-order dependence of the ground state splitting on the field strength in the perturbative regime: $\Delta E = \alpha \left[ 1 + A \operatorname{cos}(6\phi) \right]H^3$. Using the McPhase program \cite{rotter_using_2004}, we directly diagonalize the Hamiltonian~(\ref{eq:stevens_perturb}) and fit the coefficients to be $\alpha = 2.80\times 10^{-4}$ ($1.95\times 10^{-4}$)~[meV/T$^3$] and $A = 0.136$ (0.098) for CdEr$_2$Se$_4$ (CdEr$_2$S$_4$). For Dy$_2$Ti$_2$O$_7$, using the CEF parameters of Ref.~\cite{ruminy_crystal_2016}, the coefficients are calculated to be $\alpha = 2.14\times 10^{-6}$~[meV/T$^3$] and $A = 0.183$. As is compared in Fig.~\ref{fig:mono_cal} for magnetic field along the $x$ direction, the CEF ground state splittings in CdEr$_2X_4$ are indeed $\sim 10^2$ larger than that in Dy$_2$Ti$_2$O$_7$ under the same transverse magnetic field. This higher susceptibility to transverse magnetic field is a property of the full CEF Hamiltonian of CdEr$_2X_2$ as compared to that of Dy$_2$Ti$_2$O$_7$.

Our results suggest that to explain the much faster dynamics of CdEr$_2X_4$ vs Dy$_2$Ti$_2$O$_7$ only similar monopole populations combined with a much faster monopole hopping rate in the former are required, and also support the single-ion quantum tunnelling process proposed in Ref.~\cite{tomasello_single_2015} as a general monopole hopping mechanism in dipolar spin ices.  Meanwhile, it should be noted that other factors may also contribute to the high monopole hopping rates in CdEr$_2X_4$. For example, the non-vanishing components of $|J,J_z\rangle$ with $|J_z|\leq 7/2$ in  CdEr$_2X_4$ ground state doublet might induce multipolar interactions that can further increase the monopole hopping rates \cite{rau_magnitude_2015}.

In summary, neutron scattering investigations of the spin correlations in CdEr$_2X_4$ ($X$ = Se, S) confirm they are the first spinels that realize dipolar spin ice states.  High temperature Orbach behavior gives way to fast (compared to  Dy$_2$Ti$_2$O$_7$)  monopole hopping dynamics at low temperature.  Comparison of monopole populations calculated using Coulomb gas parameters estimated from the diffuse scattering experiments and bulk properties, and monopole hopping rates calculated using the CEF Hamiltonian derived from our inelastic neutron scattering data, show that the main contribution to the fast monopole dynamics of CdEr$_2X_4$ is due to the much larger hopping rate. The reproduction of the very different relaxation rates in CdEr$_2X_4$ and Dy$_2$Ti$_2$O$_7$ using realistic parameters supports the general application of this method to the description of monopole hopping processes in dipolar spin ices.

Cd$R_2X_4$ (and Mg$R_2X_4$ \cite{plessis_spin_2017}) afford new possibilities in the study of frustrated magnetism on the pyrochlore lattice, with single ion ground states~\cite{supp}, interactions, and dynamics that contrast with the well known pyrochlore oxides~\cite{gardner_magnetic_2010}. One immediate benefit of the fast monopole hopping rate in CdEr$_2X_4$ is that it enables the study of the magnetic Coulomb phase in a broader frequency regime. In particular, nonequilibrium phenomena such as the Wien effect~\cite{paulsen_far_2014, paulsen_experimental_2016}, which appear in Dy$_2$Ti$_2$O$_7$ at temperatures well below those measured here (by susceptibility), may be modified. On the other hand, if the timescale of dynamics is taken as a measure of the quantum contribution to the dynamics of a spin ice, going from slow and classical (Dy$_2$Ti$_2$O$_7$) to fast and quantum (\textit{e.g.} Tb- or Pr-based quantum spin ice candidates \cite{gingras_quantum_2014}), CdEr$_2X_4$ offer an intermediate case that may help in the extrapolation of our understanding of the former to that of the latter. 
Finally, CdEr$_2X_4$ offer the possibility to look for a new ground state of dipolar spin ice~\cite{mcclarty_chain_2015}. As is discussed in the supplemental materials~\cite{supp}, for Dy$_2$Ti$_2$O$_7$, an {\it antiferromagnetic} ordering transition at $\sim0.1$~K is expected~\cite{ melko_long_2001, mcclarty_chain_2015, ruff_finite_2005, henelius_refrustration_2016}, but is experimentally inaccessible due to the relatively high freezing temperature of $\sim0.65$~K~\cite{snyder_low_2004, pomaranski_absence_2013}. For CdEr$_2$Se$_4$, our parameters predict a {\it ferromagnetic} ordering transition at $\sim0.37$~K and a comparable freezing temperature of $\sim0.29$~K~\cite{supp}. This means that both the ferromagnetic ground state and new monopole interactions caused by the bandwidth of the spin ice states may be experimentally accessible~\cite{supp}. Further dynamical and thermodynamic measurements at low temperatures would be required to conclude whether the spin ice state that we observed at 0.07 K is an equilibrium state and to explore the possible ordering transition in CdEr$_2X_4$.

\vskip 0.1cm
% If you have acknowledgments, this puts in the proper section head.
\begin{acknowledgments}
We acknowledge valuable discussions with C. Castelnovo, M.J.P. Gingras, B. Tomasello, L. D. C.  Jaubert, H. Kadowaki, G. Chen, M. Ruminy, J. Xu, J.S. White, A. Turrini, and J.-H. Chen. We  thank V. Markushin for help with the Merlin4 cluster.
Our neutron scattering experiments were performed at the Institut Laue-Langevin ILL, Grenoble, France and the Swiss Spallation Neutron Source SINQ, Paul Scherrer Institut PSI, Villigen, Switzerland. The susceptibility measurements were carried out in the Laboratory for Scientific Developments and Novel Materials of PSI. The Monte Carlo simulations were performed on the Merlin4 cluster at PSI.
This work was supported by the Swiss National Science Foundation under Grants No. 20021-140862, No. 20020-162626, and the SCOPES project No. IZ73Z0-152734/1. S. R. G. thanks EPSRC for financial support under EP/L019760/1. V. T. and A. L. acknowledge funding from the Deutsche Forschungsgemeinschaft (DFG) via the Transregional Collaborative Research Center TRR 80. E. C. acknowledges support from the Danish Research Council for Science and Nature through DANSCATT.
\end{acknowledgments}

\bibliography{CESe}

%%%%%%%%%%%%%%%%%%%%%%%%%%%%%%%%%%%%%%%%%%%%%%%%%%%%%%%%%%

\newcommand{\ra}[1]{\renewcommand{\arraystretch}{#1}}
\newcolumntype{.}{D{.}{.}{-1}}

\renewcommand{\thefigure}{S\arabic{figure}}
\renewcommand{\thetable}{S\arabic{table}}

\renewcommand{\theequation}{\arabic{equation}}

\makeatletter
\renewcommand*{\citenumfont}[1]{S#1}
\renewcommand*{\bibnumfmt}[1]{[S#1]}
\makeatother

\setcounter{figure}{0} 
\setcounter{table}{0}
\setcounter{equation}{0} 

\onecolumngrid
\newpage
\begin{center} {\bf \large Dipolar spin ice states with fast monopole hopping rate in CdEr$\bf{_2}$\textit{X}$\bf{_4}$ (\textit{X} = Se, S) \\
 Supplementary Information} \end{center}
\vspace{0.5cm}
\twocolumngrid

%%%%%%%%%%%%%%%%%%%%%%%%%%%%%%%%%%%%%%%%%%%%%%%%%%
 \subsection{Sample preparation}
 
The polycrystalline samples of CdEr$_2X_4$ ($X=$ Se, S) were prepared by solid state synthesis from binary Er and Cd selenides and sulfides. The binary Cd$X$ were synthesized from the elemental Cd-114, while the Er$_2X_3$ were prepared from the high purity Er chips (99.9 \%, Chempur) and elemental S (99.999 \%, Strem Chemicals) or Se (99.999 \%, Alfa Aesar). Selenium was additionally purified by zone melting. To reduce the oxide impurity Er$_2$O$_2X$ which easily forms in the open air, all preparation procedures (quartz ampoule filling, reacted mixture regrinding, and pellets pressing) were performed in an argon box with an O$_2$ and H$_2$O content of $\sim 1$ ppm. To reach full homogeneity, at least three sintering cycles of synthesis of the binary Er and Cd chalcogenides were performed. The phase purity of the binary compounds was checked by x-ray powder diffraction. Finally, the ternary $^{114}$CdEr$_2X_4$ were prepared by two consecutive synthesis at 800 $^\circ$C for one week each.

The single crystals of CdEr$_2X_4$ were grown by the chemical transport reactions method. As starting materials the preliminary synthesized polycrystalline powders were used. For the growth, several transport agents were probed, including chlorine, bromine and iodine. We found that only the iodine is suitable for the growth of the ternary phase, while in the case of chlorine or bromine the final product contained mainly binary Cd and Er chalcogenides. The growth process was performed in a two-zone furnace with the hot part temperature of 950 $^\circ$C and a temperature gradient of about 40 $^\circ$C. The time for one crystal growth experiment was between 1 and 1.5 months. As a result, the octahedron-like single crystals with dimension up to 1.5 mm of the edge were obtained.

\subsection{Sample characterizations and impurities in CdEr$_2$Se$_4$}

\begin{figure} [h!]
\includegraphics[width=0.40\textwidth]{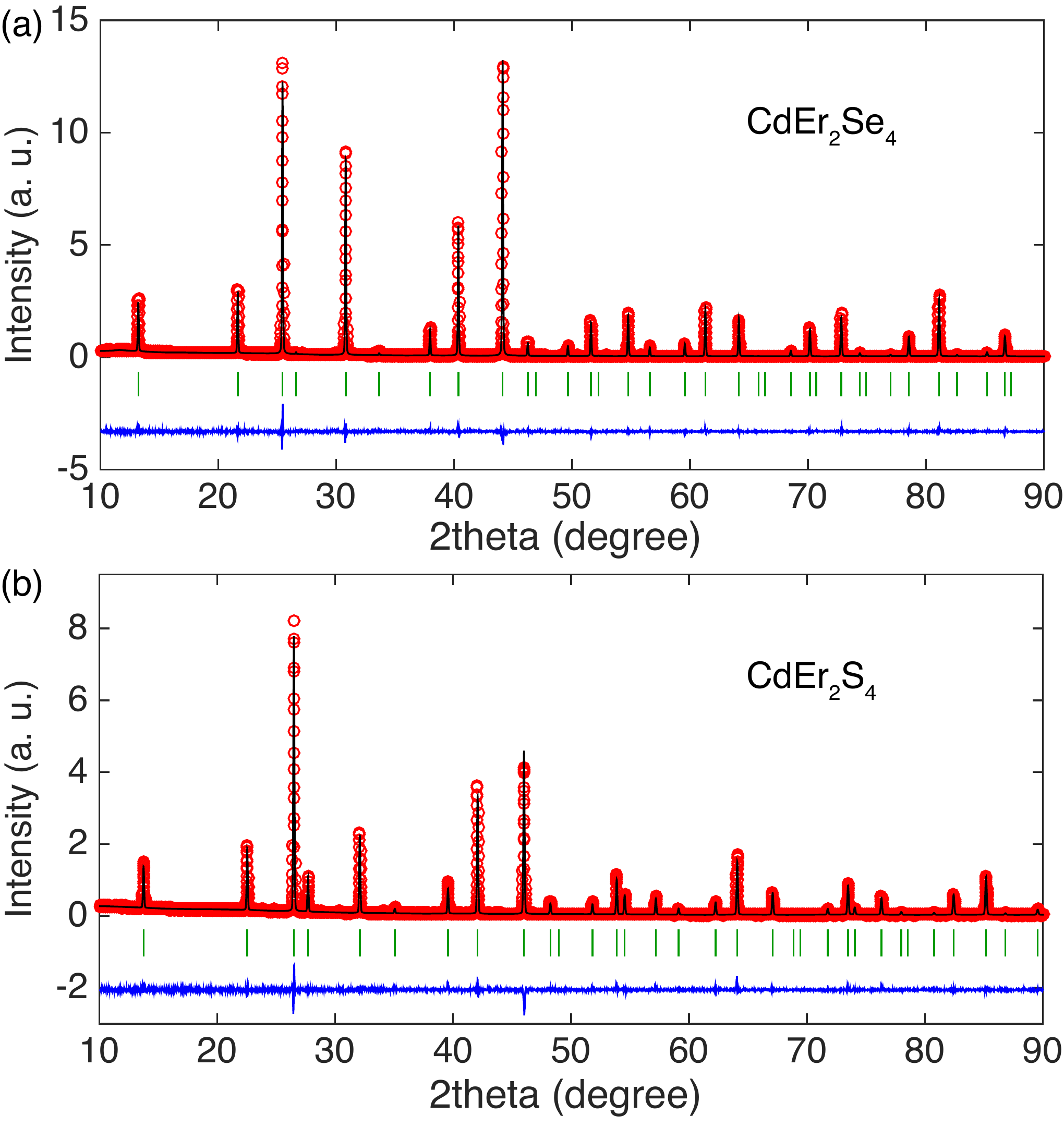}
\caption{Refinement results of the X-ray diffraction data measured at room temperature for (a) $^{114}$CdEr$_2$Se$_4$ polycrystalline sample and (b) $^{114}$CdEr$_2$S$_4$ polycrystalline sample. Data points are shown as red circles. The calculated pattern is shown as black solid line. The vertical bars mark the positions of Bragg peaks. And the blue line at the bottom shows the difference of data and calculated intensities.}
\label{fig:xrd}
\end{figure}

The purity content and crystal structure of the samples were checked by conventional X-ray powder diffraction on polycrystalline samples and crashed single crystals. Fig.~\ref{fig:xrd} shows the refinement results for the $^{114}$CdEr$_2$Se$_4$ and $^{114}$CdEr$_2$S$_4$ polycrystalline samples using cubic $Fd\overline{3}m$ symmetry expected for the normal spinel structure. Tab~\ref{tab:xrd} lists the refined size of the unit cell, fractional position for the Se or S ions, and the goodness-of-fit parameters. No inversion between Cd and Er can be observed. No peaks from impurities are detectable, implying their tiny amount. However, at low temperatures, weak magnetic Bragg peaks of the Er$_x$Se$_y$ impurities~\cite{calder_neutron_2010s} are discernible in the neutron diffuse scattering experiment shown in Fig. 2 of the main text. 

\begin{figure}
\includegraphics[width=0.4\textwidth]{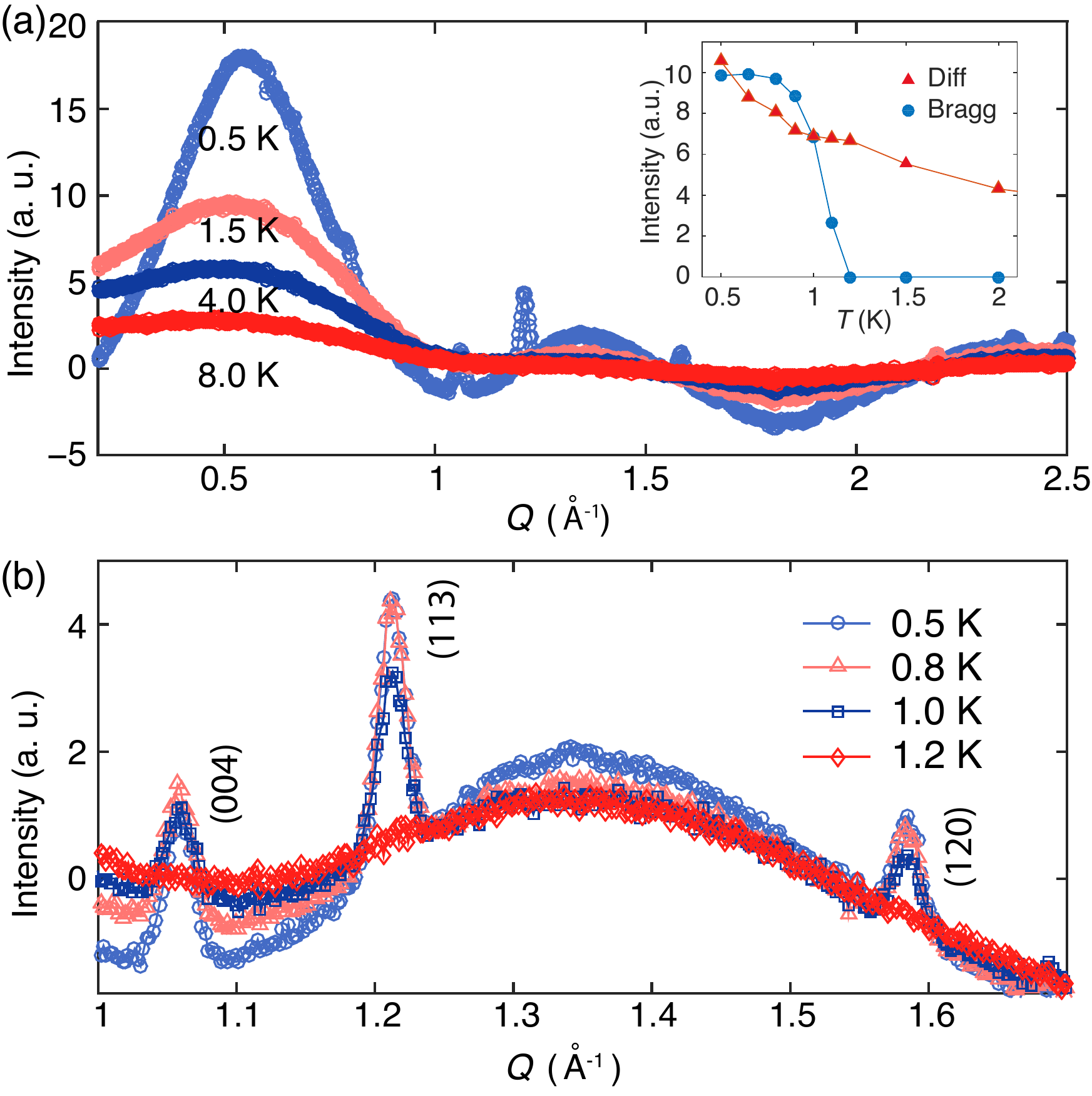}
\caption{(a) Non-polarized neutron diffraction results for CdEr$_2$Se$_4$ measured on D20 at 0.5, 1.5, 4, and 8 K. The inset compares the temperature dependence of the Bragg peak intensity at 1.22 \AA$^{-1}$ and the integrated diffuse scattering intensity in $1.325 < Q < 1.375$ \AA$^{-1}$. (b) Detailed temperature dependence of the intensities in the region of 1 $\sim$ 1.7 \AA$^{-1}$ in between 0.5 and 1.2 K. The Bragg peaks are indexed for the impurity of Er$_2$Se$_3$.}
\label{fig:impurity}
\end{figure}

The extrinsic origin of the weak Bragg peaks is evident in their different temperature dependence compared with the broad diffuse scattering. Fig. \ref{fig:impurity} shows the non-polarized neutron diffraction results measured on D20 at ILL with the 20~K measurement subtracted as the background. The setup with 2.41 \AA\ incoming neutron wavelength was employed. As can be seen in the inset of Fig. \ref{fig:impurity}a and Fig. \ref{fig:impurity}b, intensities of the sharp peaks saturate at temperatures below 0.8 K while the broad peaks from the diffuse scattering continue their growth, evidencing their different origins. 

%--------------------------------------------------------
%--------------------------------------------------------
%--------------------------------------------------------
\begin{table} [h]
\caption{Refinement results of the X-ray diffraction data for CdEr$_2$Se$_4$ and CdEr$_2$S$_4$. The listed parameters are the refined size of the unit cell $a$, fractional position $x$ of the $X=$ Se or S ions, and the goodness-of-fit $R_\textrm{p}$, $R_\textrm{wp}$, and $\chi^2$.}
\label{tab:xrd}
\centering
\ra{1.2}
\begin{tabular}{lccccc}
\toprule
 & $a$ (\AA) & $X$ ($x$) & $R_\textrm{p}$ & $R_{\textrm{wp}}$ & $\chi^2$ \\
 \hline
CdEr$_2$Se$_4$	& 11.6097(1) & 0.2566(1) & 13.0  & 14.6 & 2.16 \\
CdEr$_2$S$_4$	& 11.1527(1) & 0.2589(2) & 19.7  & 19.1 & 1.89\\
\botrule
\end{tabular}
\end{table}
%--------------------------------------------------------

%--------------------------------------------------------
%--------------------------------------------------------

\subsection{CEF levels for CdEr$_2$Se$_4$ and CdEr$_2$S$_4$}
Table \ref{tab:cef} lists the crystal-electric-field (CEF) levels for CdEr$_2$Se$_4$ and CdEr$_2$S$_4$ fitted from inelastic neutron scattering experiments. The CEF levels for CdEr$_2$Se$_4$ fitted from magnetization measurements \cite{lago_cder2se4_2010s} are also listed, showing strong deviations at high energies.

%--------------------------------------------------------
\begin{table} [h]
\caption{CEF levels for CdEr$_2$Se$_4$ (column 2) and CdEr$_2$S$_4$ (column 3) fitted from inelastic neutron scattering experiments. CEF levels for CdEr$_2$Se$_4$ fitted from magnetization measurements from Ref. \cite{lago_cder2se4_2010} are listed in column 1. Energy unit is meV.}
\label{tab:cef}
\centering
\ra{1.2}
\begin{tabular}{cccc}
\toprule
 & CdEr$_2$Se$_4$ \cite{lago_cder2se4_2010} & CdEr$_2$Se$_4$ & CdEr$_2$S$_4$  \\
 \hline
$\Gamma_5^+ \oplus \Gamma_6^+$	& 0  & 0 & 0 \\
$\Gamma_4^+$	& 3.64  & 3.95  & 5.08\\
$\Gamma_4^+$	& 7.22  & 5.65  & 6.91\\
$\Gamma_5^+ \oplus \Gamma_6^+$	& 7.65  & 8.93 & 10.47 \\
$\Gamma_4^+$	& 8.70  & 9.76  & 11.35\\
$\Gamma_4^+$	& 21.70  & 26.28  & 30.46\\
$\Gamma_4^+$	& 23.31  & 29.11  & 33.78\\
$\Gamma_5^+ \oplus \Gamma_6^+$	& 23.81  & 29.21 & 34.00 \\
\botrule
\end{tabular}
\end{table}
%--------------------------------------------------------
%sele_lago= 0, 3.645, 7.216, 7.650, 8.704, 21.700, 23.312, 23.8080

%cef_sele = [0, 3.95, 5.65, 8.93,  9.76, 26.28, 29.11, 29.21];

% sulf
%cef_sulf = [0, 5.08, 6.91, 10.47, 11.35, 30.46, 33.78, 34.00];

\subsection{Scaled CEF levels for CdRE$_2$Se$_4$}
With the classical spin ice state established in CdEr$_2X_4$ ($X=$ S, Se), it is tempting to look for the quantum spin ice state in the CdRE$_2X_4$ series. Until now, four other compounds with RE = Dy, Ho, Tm, and Yb have been successfully synthesized \cite{lau_geometrical_2005s, yaouanc_evidence_2015s, yoshizawa_high_2015s, higo_frustrated_2016s}. Their CEF parameters can be approximated using the scaled values of CdEr$_2$Se$_4$ \cite{bertin_crystal_2012s}: 
\begin{equation}
A_l^m(R')=\frac{a^{l+1}(R)}{a^{l+1}(R')}A_l^m(R)\ \text{,}
\end{equation}
where $R$ and $R'$ represent different rare-earth ions, $a$ is the size of the unit cell and the values for the CdRE$_2$Se$_4$ compounds are listed in Tab. \ref{tab:cef_scale}, and $A_l^m$ are the Hutchings CEF parameters that can be transformed from the Stevens CEF parameters $B_l^m$ following the relation:
\begin{equation}
B_l^m = A_l^m\langle r^l \rangle \theta_l \text{.}
\end{equation}
In this expression, $\langle r^l \rangle$ is the expectation value of the $l$-th power of the $f$-electron radius and can be found in Ref. \cite{freeman_dirac_1979s}, $\theta_l$ is the Stevens factor that can be found in Ref. \cite{jensen_rare_1991s}.
\begin{table}[t!]
\caption[CdRE$_2$Se$_4$ CEF parameters]{The Wybourne CEF parameters for the CdRE$_2$Se$_4$ compounds scaled from the refined CdEr$_2$Se$_4$ values. The sizes of the unit cell used in the scaling calculations are also listed. }
\label{tab:cef_scale}
\centering
\ra{1.2}
\small
\begin{tabular}{c . . . .}
\toprule
 & \multicolumn{1}{c}{CdDy$_2$Se$_4$} & \multicolumn{1}{c}{CdHo$_2$Se$_4$} & \multicolumn{1}{c}{CdTm$_2$Se$_4$} & \multicolumn{1}{c}{CdYb$_2$Se$_4$} \\
\hline
unit cell (\AA) & \multicolumn{1}{c}{11.467}  & \multicolumn{1}{c}{11.638} & \multicolumn{1}{c}{11.560} & \multicolumn{1}{c}{11.528} \\
\hline
\multicolumn{5}{c}{Wybourne CEF parameters (meV)}\\
\hline
          $L_2^0$ &  -27.88    &  -26.59    &  -24.82   &  -23.96 \\
          $L_4^0$ & -124.91    & -114.88    & -100.99   &  -95.04 \\
          $L_4^3$ & -113.33    & -104.23    &  -91.63   &  -86.23 \\
          $L_6^0$ &   30.68    &   27.70    &   23.28   &   21.53 \\
          $L_6^3$ &  -23.10    &  -20.85    &  -17.53   &  -16.21 \\
          $L_6^6$ &   11.53    &   10.40    &    8.75   &    8.09 \\
\botrule 
\end{tabular}
\end{table}
%--------------------------------------------------------
\begin{figure} [t]
\centering
\includegraphics[width=0.42\textwidth]{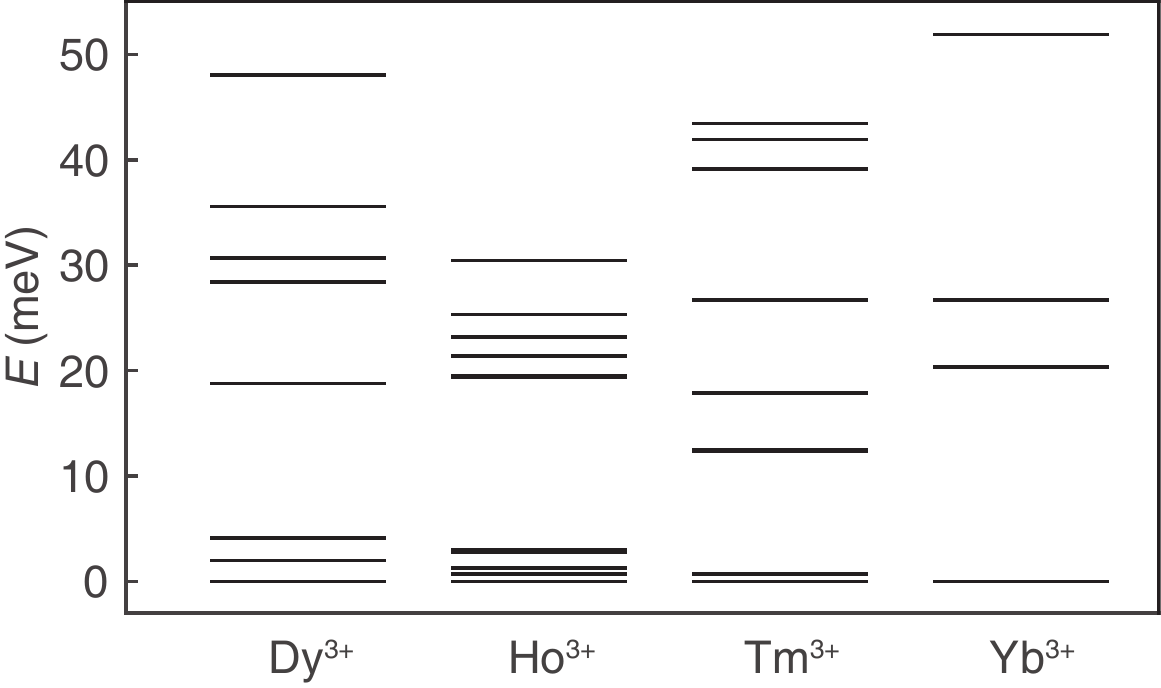}
\caption{CEF levels for CdRE$_2$Se$_4$ with RE = Dy, Ho, Tm, and Yb using the CEF parameters listed in Tab. \ref{tab:cef_scale}. }
\label{fig:scale_level}
\end{figure}
%--------------------------------------------------------

Tab. \ref{tab:cef_scale} lists the scaled Wybourne CEF parameters for the CdRE$_2$Se$_4$ compounds. The calculated CEF levels are shown in Fig. \ref{fig:scale_level}, and the ground states are:
\begin{align}
\label{eq:scale_ground}
\text{Dy}^{3+}\ |\pm\rangle =\ &0.176 |15/2, \pm 13/2\rangle
              \mp   0.350 |15/2, \pm 7/2\rangle \nonumber \\
              + &0.273 |15/2, \pm 1/2\rangle 
              \pm   0.554 |15/2, \mp 5/2\rangle \nonumber \\ 
              + &0.682 |15/2, \mp 11/2\rangle \text{,} \nonumber \\               
\text{Ho}^{3+}\ |\pm\rangle =\ & 0.157 |8, \pm 8\rangle
              \mp   0.456 |8, \pm 5\rangle \nonumber \\
              + &0.252 |8, \pm 2\rangle
               \pm   0.096 |8, \mp 1\rangle \nonumber \\ 
              + &0.743 |8, \mp 4\rangle 
              \pm 0.377 |8, \mp 7\rangle \text{,} \nonumber \\ 
\text{Tm}^{3+}\ |\phi\rangle =\ & 0.659 |6, \pm 6\rangle
                \pm   0.200 |6, \pm 3\rangle
              + 0.2235 |6, 0\rangle \text{,} \nonumber \\
\text{Yb}^{3+}\ |\pm \rangle =\ &-0.256 |3.5, \pm 3.5\rangle
                \pm   0.389 |3.5, \pm 0.5\rangle \nonumber \\
               &+ 0.885 |3.5, \mp 2.5\rangle \text{.}            
\end{align}

Firstly, it is clear that the ground state of Tm$^{3+}$ is a singlet due to its non-Kramers character, while for spin ice, a doublet ground state is required. For the remaining compounds where a doublet CEF ground state is realized, spins in CdDy$_2$Se$_4$ and CdYb$_2$Se$_4$ exhibit Heisenberg-like character, with $g$-factors of $g_{\perp} = 5.69$, $g_{\parallel}=6.38$ for Dy$^{3+}$ and $g_{\perp} = 2.16$, $g_{\parallel}=3.97$ for Yb$^{3+}$. For the Ho$^{3+}$ spin, an Ising 
character with $g_{\perp} = 0$, $g_{\parallel}=4.43$ is observed, which satisfies the local Ising condition to realize the spin ice state. Therefore, from the CEF point of view, CdHo$_2$Se$_4$ might be the most promising compound to realize the quantum spin liquid state. However, it should be noted that the lowest CEF excited states in CdDy$_2$Se$_4$, CdTm$_2$Se$_4$, and CdHo$_2$Se$_4$ are lying at energies below $\sim 2$ meV. Such low-lying excited levels might renormalize the spin couplings and make our single-ion analysis inappropriate \cite{molavian_dynamically_2007s, yaouanc_evidence_2015s}.

%\vspace{0.5cm}
\subsection{Dipolar spin ice state in CdEr$_2$S$_4$}
 
\begin{figure}
\includegraphics[width=0.42\textwidth]{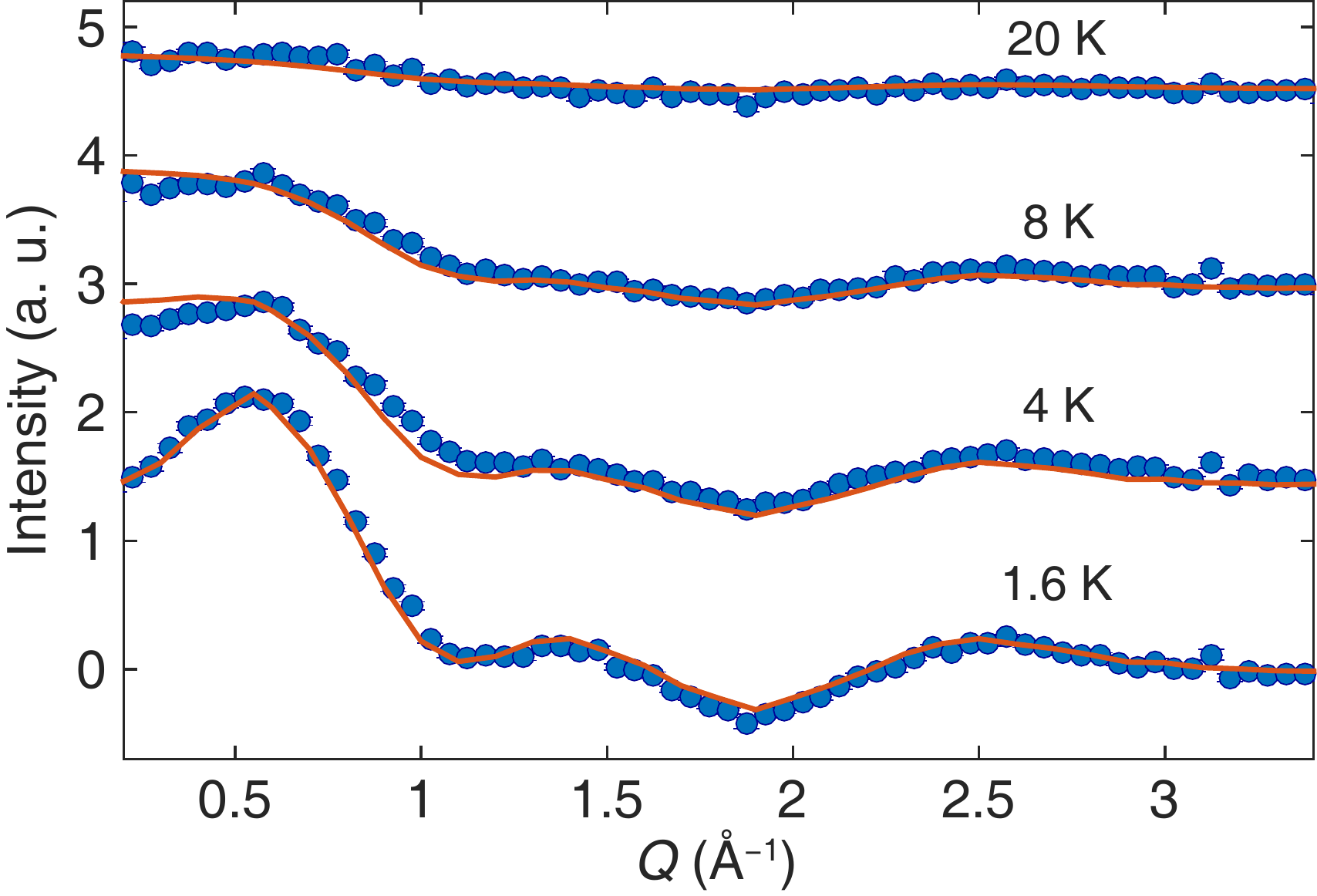}
\caption{Non-polarized neutron diffuse scattering results for CdEr$_2$S$_4$ measured at 1.6, 4, 8, and 20 K with the 50 K data subtracted as the background. The data at 4, 8, and 20 K are shifted by 1.5, 3.0, and 4.5 along the $y$ axis, respectively. The mean-field calculation results assuming only dipolar interactions are shown as the solid lines.}
\label{fig:diff_sulf}
\end{figure}

Ice-correlations similar to that of CdEr$_2$Se$_4$ are also observed in CdEr$_2$S$_4$. Fig. \ref{fig:diff_sulf} presents the non-polarized neutron diffuse scattering results for CdEr$_2$S$_4$ measured on DMC at PSI with the setup of 2.46 \AA\ incoming neutron wavelength. The 50~K measurement has been subtracted as the background. Similar to the CdEr$_2$Se$_4$ results shown in Fig.~2 of the main text, broad peaks at $\sim 0.6$, 1.4, and 2.5 \AA$^{-1}$ are observed at low temperatures, which suggest similar ice-correlations in CdEr$_2$S$_4$.

Mean-field calculations were performed to confirm the ice-correlations in CdEr$_2$S$_4$ \cite{kadowaki_neutron_2002s}. Denoting the $\alpha$ component ($\alpha = x,y,z$) of the $\nu$-th unit-length spin ($\nu = 1,2,3,4$) in the $n$-th primitive unit cell as $S_{n,\nu, \alpha}$, the Hamiltonian on the pyrochlore lattice can be explicitly expressed as:

\begin{widetext}
\begin{align}
\label{eq:ces_mf}
\mathcal{H} = &- E_a \sum_{n, \nu} 
\left[ 
\left( \hat{n}_{\nu} \cdot  S_{n, \nu} \right)^{2} 
- |S_{n, \nu} |^{2}
\right] 
- J_{1} \sum_{\langle n, \nu ; n', \nu' \rangle} 
S_{n, \nu} \cdot S_{n', \nu'} 
\nonumber \\
&+ Dr_0^3 \sum_{\langle n, \nu ; n', \nu' \rangle}
\Bigl[
\frac{S_{n, \nu} \cdot  S_{n', \nu'} }
{ |r_{n, \nu ; n', \nu'}|^3 } 
\nonumber 
-\frac{3 
\left( S_{n, \nu} \cdot r_{n, \nu ; n', \nu'} \right)
\left( S_{n', \nu'} \cdot r_{n, \nu ; n', \nu'} \right)
}
{ | r_{n, \nu ; n', \nu' } |^5 }
\Bigr] \\
= &- \sum_{n, \nu, \alpha, n', \nu', \beta} J_{n, \nu, \alpha; n', \nu', \beta} S_{n, \nu, \alpha} S_{n', \nu', \beta}\ \text{,}
\end{align}
\end{widetext}
where the $E_a$ term represents the easy-axis anisotropy and $\hat{n}_{\nu}$ is the unit vector along the easy axis of the $\nu$-th spin. Fourier transform of the real-space coupling $J_{n, \nu, \alpha; n', \nu', \beta}$ leads to the $12\times12$ coupling matrix $J_{k; \nu, \alpha; \nu', \beta}$ in reciprocal space, which is then diagonalized:
\begin{equation}
\label{eq:ces_mf_eigen}
\sum_{\nu', \beta} J_{k; \nu, \alpha; \nu', \beta}
u_{k; \nu', \beta}^{(\rho)}
 = 
\lambda_{k}^{(\rho)} u_{k; \nu, \alpha}^{(\rho)}  \;,
\end{equation}
where $\lambda_{k}^{(\rho)}$ with $\rho = 1, 2, ..., 12$ denotes the eigenvalues and $u_{k}^{(\rho)}$ denotes the corresponding eigenvectors. The global maximum of $\lambda_{k}^{(\rho)}$ determines the long-range order transition under the mean-field approximation, with the transition temperature $T_c$ as:
\begin{align}
k_{\rm B} T_{\rm c}  &= 
\frac{2}{3} 
[\lambda_k^{(\rho)}]_{{\rm max}}  \notag 
\end{align}
%--------------------------------------------------------
%--------------------------------------------------------

The paramagnetic susceptibility at $T_{\textrm{MF}} > T_c$ can be approximated by the eigenvalues and eigenfunctions:
\begin{equation}
\chi_{k; \nu, \alpha; \nu', \beta}
=
\frac{N \mu^{2}}{V} 
\sum_{\rho} 
\frac{ 
u_{k; \nu, \alpha}^{(\rho)} 
u_{k; \nu', \beta}^{(\rho) *}
}
{
3 k_{\rm B} T_{\rm MF} - 2 \lambda_k^{(\rho)}
}
\;,
\end{equation}
where $N$ is the total number of the unit cell, $V$ is the volume of the system, and $\mu$ is the size of the magnetic moment. The cross section of the magnetic scattering can be expressed as:
\begin{align}
\frac{d \sigma}{d \Omega} &
\bigl( Q = \tau + k \bigr)
= P
 f(Q)^{2} k_{\rm B} T
\sum_{\alpha, \beta, \nu, \nu'}
\left( 
\delta_{\alpha \beta} - \hat{Q}_{\alpha} \hat{Q}_{\beta} \right) \notag \\
&\times
\chi_{k; \nu, \alpha; \nu', \beta}
\cos \left[ \tau \cdot 
\left( r_{\nu} - r_{\nu'} \right) \right]- K f(Q)^2
\; ,
\label{eq:mf}
\end{align}
where $P$ is a constant, $f(Q)$ is the magnetic form factor, $\tau$ is the reciprocal lattice vector, and $r_{\nu}$ denotes the position of the $\nu$-th atom in the first primitive cell. The additional term of $Kf(Q)^2$ accounts for the subtracted spin correlations at 50 K.

%--------------------------------------------------------
\begin{figure} [h]
\centering
\includegraphics[width=0.42\textwidth]{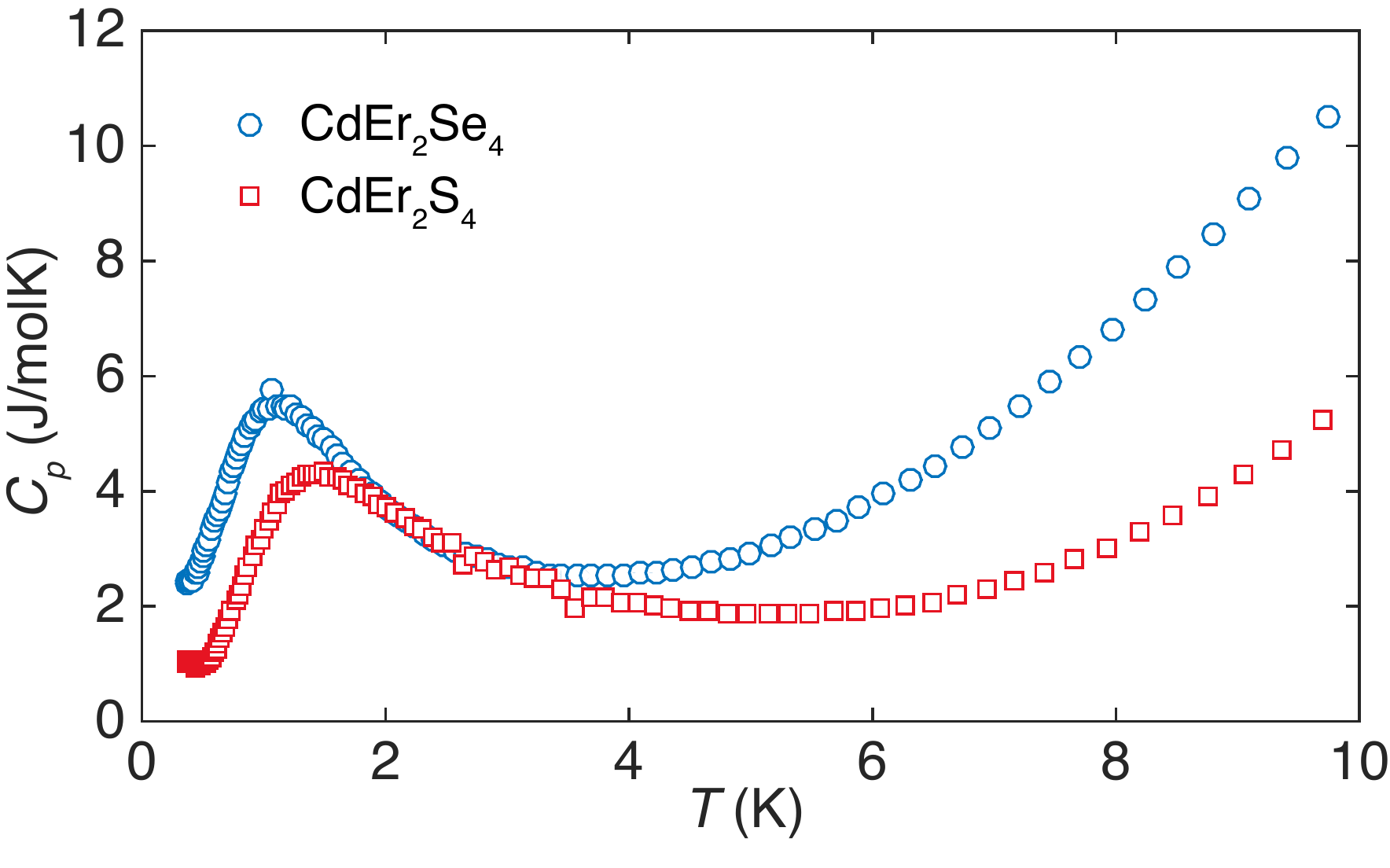}
\caption{Specific heat for CdEr$_2$Se$_4$ and CdEr$_2$S$_4$ single crystals. Tails at the lowest temperature might be due to impurities.}
\label{fig:cp}
\end{figure}
%--------------------------------------------------------

To account for the Ising character of the Er$^{3+}$ spin, a high anisotropy of $E_a=100$ K was used. The dipolar interactions with $D = 0.69$ K was truncated beyond the length of 5 unit cells. Since $T_c$ is an effective temperature that can be different from the real ordering temperature, the mean-field temperature $T_{\textrm{MF}}$ was used as a fitting parameter \cite{kadowaki_neutron_2002s}. Fig. \ref{fig:diff_sulf} presents the fitted results with $J_1 = 0$. In this case, $T_c$ is calculated to be 1.8 K, and the fitted $T_{\textrm{MF}}= 1.9$, 2.3, 3.8, and 15.8 K for the data measured at 1.6, 4, 8, and 20 K, respectively. Thus the ice-correlation is proved to exist in CdEr$_2$S$_4$. 

In our mean-field calculation, the variance of $J_1$ does not affect the goodness-of-fit as long as $J_1/3+5D/3>0$. To obtain the monopole chemical potential, we measured the specific heat $C_p$ for CdEr$_2$S$_4$ single crystals. As is shown in Fig. \ref{fig:cp}, the $C_p(T)$ maximum of CdEr$_2$S$_4$ is at $\sim1.4$ K, which enable us to fix the monopole chemical potential to 3.84 K in CdEr$_2$S$_4$ (see main text).

\subsection{Freezing and ordering temperatures in CdEr$_2$Se$_4$}

The true ground state of a dipolar spin ice has again become a topical question~\cite{pomaranski_absence_2013s}.  An ordering transition is expected due to the bandwidth of the dipolar spin ice microstates~\cite{melko_long_2001s},  the transition temperature and eventual ground state being controlled by $D/J_2$~\cite{ruff_finite_2005s,McClarty:2014vcs,henelius_refrustration_2016s}.  In Dy$_2$Ti$_2$O$_7$ a transition to {\it antiferromagnetic} order is expected at $\sim0.1$~K~\cite{yavorskii_dy2ti2o7_2008s,McClarty:2014vcs,henelius_refrustration_2016s}, far below temperatures at which equilibration is easily possible, \textit{i.e.} 0.65~K~\cite{snyder_low_2004s}. 
 
The freezing temperature of 0.29 K for CdEr$_2$Se$_4$ is estimated from the 0.65~K freezing temperature of Dy$_2$Ti$_2$O$_7$. With the known monopole parameters, the temperature dependence of the monopole density $\rho(T)$ can be obtained using the Debye-H\"uckel theory \cite{zhou_high_2011s}. Assuming the monopole hopping rate $u$ to be temperature independent and $u\rho$ to be the same at the freezing temperature for CdEr$_2$Se$_4$ and Dy$_2$Ti$_2$O$_7$, we first calculate the monopole density in Dy$_2$Ti$_2$O$_7$ at 0.65~K, then divide by 100 to account for the increased value of $u$ in CdEr$_2$Se$_4$, and finally locate the freezing temperature in CdEr$_2$Se$_4$ by following its $\rho(T)$ relation using Debye-H\"uckel theory. 

The ordering temperature $T_c=0.37$~K for CdEr$_2$Se$_4$ is estimated from the $J_2/D$-$T_c/D$ phase diagram in Ref.~\cite{mcclarty_chain_2015s}. Considering the $-3$ times difference in the definition of $J_2$, the ratio $J_2/D$ is found to be $-0.2$, which leads to $T_c = 0.37~K$ under a linear extrapolation of the boundary between the classical spin ice and ferromagnetically long-range ordered phases. This means that in CdEr$_2$Se$_4$ there is a chance not only to investigate a spin ice with an alternative ground state; but also to study the effect on monopole dynamics of reaching a temperature comparable to the bandwidth of the spin ice states where new monopole-monopole interaction terms may appear. 

\subsection{Note for the ground state doublet splitting in Dy$_2$Ti$_2$O$_7$}
Private communications with the authors of Ref. \cite{tomasello_single_2015s} confirm a typo for the fitted parameter $\alpha$ in Eq.~(9) of their publication. The correct value should be $\alpha = 3.29 \times 10^{-6}$~[meV/T$^3$], which we reproduce using the same CEF parameters. The similar but different values of $\alpha = 2.14\times 10^{-6}$~[meV/T$^3$] and $A = 0.183$ reported in our main text are calculated with refined CEF parameters~\cite{ruminy_crystal_2016s}, which were not available to the authors of Ref.~\cite{tomasello_single_2015s}.

\end{document}